\documentclass[aps,onecolumn,prd,nofootinbib]{revtex4}
\usepackage{amsmath}
\usepackage{graphicx}
\usepackage{dcolumn}
\usepackage{bm}
\usepackage{amssymb}
\usepackage{latexsym}

\begin{document}

\title{Inspiraling Corrugation-Induced Quantum Effects on Neutron Star Binary Plane}

\author{Jing Wang\footnote{Email address: joanwangj@mailbox.gxnu.edu.cn}}
\affiliation{School of Physical Science and Technology, Guangxi Normal University,
   Guilin, 541004, P. R. China}

\begin{abstract}
We use the path-integral formula and investigate some dynamical quantum effects induced by the inspiraling lateral corrugation of orbital plane in gravitationally bound neutron star (NS) binaries, with orbital separation of $10^9$ m. Based on Dewitt's approach, we calculate the gravitational Casimir energy cost of the binary plane, which consists of statically gravitational effects and deformation-induced effects. It is found that the static effects include a term coming from the self-gravity of the orbital plane and the contribution of Newtonian gravitational potential of the binary system. While the deformation-induced effect also results from two parts, i.e. the instability of orbital binding energy, scaling as $\frac{1}{(R-r)^2}$, and the dynamically Casimir energy cost of the orbital binding energy, decaying as $\frac{1}{(R-r)^4}$. The dynamically gravitational Casimir phenomena and the corresponding energy cost modify the spiral-in orbital motion of the binary and thus the frequency of released gravitational waves (GWs). We consider the mechanical response of two NS components and qualitatively study the corrections to the orbital motion of the system and the GW frequencies. It is found that the dynamical Casimir effects exert a dissipative force on the binary plane, depending on the frequency of GWs. The resultant dissipation may enhance with the decaying separation and increasing GW frequencies, which subsequently accelerates the orbital decay of the binary. However, the dissipation rate just has an order of $10^{-70}$ eV/s. So the corrections to the dynamics of NS binaries are very marginal, by considering the wide separation, the cosmological coalescence time, and low-frequency GWs of the system.
\end{abstract}

\maketitle

\section{Introduction}

The Casimir effect \cite{Plunien:1986ca, Mostepaneko:1997book} demonstrates the nontrivial properties of quantum fluctuations of the vacuum energy density, which arises from the disturbance of the vacuum of electromagnetic field subject to the boundary conditions imposed by two conducting plates \cite{Casimir1948}. Even though very large distance between two macroscopic bodies on which the virtual photon emitted from one body cannot reach the other one during its lifetime, the correlated oscillations of the quantized electromagnetic field in the vacuum state appears nontrivial, which result in a non-neglectable Casimir force \cite{Bordag:2001qi} and give rise to considerable effects and influences on the macroscopic systems. The Casimir force provides a direct investigation for the quantum effects on the systems consisting of macroscopic objects on various scales, e.g. \cite{Mostepanenko:1988bs, Krech1994book}. Correspondingly, the quantum fluctuations of different fields are allowed to be learned, by measuring the quantum force between macroscopic bodies. In addition to the standard boundary conditions coming from some simple geometries, e.g. two parallel plates or perfect spheres, the investigations for Casimir force were generalized to both deformations of random surfaces \cite{Balian:1977qr, Bordag1995, Li:1991, Li:1992} and moving boundaries, which refers to as the dynamical Casimir effects \cite{Moore:1970, Fulling:1976, Ford:1982ct, Jaekel:1992, MaiaNeto:1993zz, Meplan:1996zz, Lambrecht:1996un, Haro:2006zz, Dodonov:2010zza, RubioLopez:2016wme, Akopyan:2020xqu} (see \cite{Dodonov:2020eto} for a recent review). In order to study the dynamical Casimir effects, several relevant approaches are developed and employed to solve various cases. For example, a multiple scattering approach \cite{Balian:1977qr} is applied to compute the interactions for arbitrary geometry, with curvature, in a perturbation series. When dealing with the small deviations from plane parallel geometry, it \cite{Bordag1995} is treated as an additive summation of van der Waals-type pairwise potentials. Both methods, combining with the perturbative approach \cite{Ford:1982ct}, suffer from cumbersome treatments of the boundary conditions. However, a path-integral approach \cite{Li:1991, Li:1992}, which was originally developed for the study of thermal fluctuations, can easily implement the boundary conditions and perturbatively compute the corrections to the deformations, when it is adapted to study quantum fluctuations. It has been recognized that a dissipative radiation reaction force arises from damped motion of scatterers because of energy conservation, if the fields are perturbed by some types of motion associate with production of photons \cite{MaiaNeto:1993zz}. When dealing with the creation and annihilation of particles due to the modifications to the structure of quantum vacuum by moving boundaries, different situations have been calculated, such as fast oscillations \cite{Ji:1997gh, Crocce:2002hd} and rotating wave approximation \cite{Schaller:2002yjj}. A Hamiltonian approach \cite{Haro:2006zz} also has been introduced to calculate the radiation-reaction force during the motion.

Because of the collective nature, Casimir effects strongly depend on geometry of objects or boundaries, which has been designed experimentally with improved techniques for investigation \cite{Roy:1999zz, Chen:2002zzb}. Since the fluctuation-induced interactions between manifolds or objects immersed in a correlated medium have been computed by a general field-theoretical approach \cite{Li:1991, Li:1992}, the perturbative path-integral quantization techniques \cite{Golestanian:1997ks, Golestanian:1998bx} have been used to calculate the slightly arbitrary deformations in space and time, which subject to boundary conditions and display strong corrections to the proximity approximation \cite{Emig:2001dx, Emig:2002qpp}. The effects caused by short wavelengths were calculated by an approximation scheme based on geometric optics \cite{Jaffe:2003mb}, which can be generalized to many other situations with different geometries and boundary conditions. When calculating the periodically deformed configurations, a novel approach can non-perturbatively treat the disconnected objects \cite{Emig:2002xz}. Based also on the path-integral approach \cite{Li:1991, Li:1992}, a non-perturbative method \cite{Buscher:2004tb} was proposed to study the periodically deformed surfaces with both small surface curvatures and strong deformations, even edges. Various closed time-dependent geometries with different boundary conditions also were considered \cite{Dalvit:2006ir}.

It was found that macroscopic objects with given boundary conditions cause nontrivial gravitational Casimir effect \cite{Panella:1993pi}, which modifies the long-range gravitational interactions by imposing an interaction potential. The gravitational interaction causes a small correction to the Casimir energy of a massless scalar field in a rigid Casimir cavity in a slightly curved, static spacetime background \cite{Sorge:2005ed, Sorge:2019ldb}. When considering the non-relativistic rotation of a pair of massive moving walls, there is no first-order gravitomagnetic effect in the vacuum energy shift yet \cite{Sorge:2009zz}. In some wide neutron star (NS) binaries with separation of $R\sim10^9$ m, which may coalesce on the cosmological time, the two NS components orbit with each other on the binary plane. Subject to gravitational interactions, the system undergoes spiral-in orbital motion on the binary plane, which periodically corrugates the boundary of the orbital plane. During the inspiraling process, the binary loses orbital binding energy and radiates gravitational waves (GWs), or spin-2 gravitons. The orbital rotation of NS binary gives rise to a gravitoelectricmagnetic field, analogy with the electromagnetism, which leads to small corrections to the gravitational Casimir energy, when the binary decays a nontrivial orbital separation $r$ after several periods. Consequently, a gravitational Casimir force per area \cite{wang2021} appears,
\begin{equation}
F_{\rm Cas}\propto-\frac{\pi^2\hbar c}{240(2r)^4},
\end{equation}
which provides a direct link between the gravitationally bound relativistic NS binary and quantum effects. By taking the cosmological coalescence time into account, the two massive stars orbit with each other and move closer and closer, forming a slowly corrugating boundary of the binary plane. It was found that a slowly moving boundary is imposed by a frictional force \cite{Dodonov1989vv},
\begin{equation}
F_{\rm friction}\propto F_{\rm Cas}(\frac{\dot{r}}{c})^2,\label{friction}
\end{equation}
which is responsible for a dissipation mechanism of the system, and $c$ therein is the light speed. In this paper, we adopt the path-integral approach \cite{Li:1991, Li:1992} and investigate the gravitational Casimir effects on the NS binary plane with lateral corrugations arising from the gravitation-induced spiral-in orbital motion. With the assumption of weak-field approximation to general relativity, the spiral-in dynamics of wide NS binaries can be described by Maxwell-like formulation of the linearized Einstein field equations, i.e. the gravitoelectromagnetic field equations \cite{Mashhoon:1999nr}. Both the gravitoelectric and gravitomagnetic fields should have the form of transverse plane waves,
\begin{equation}
h^{\rm E}_{ij}=\mathcal{E}_{ij}e^{i(\vec{k}\cdot\vec{r}-\omega t)},~~h^{\rm M}_{ij}=\mathcal{M}_{ij}e^{i(\vec{k}\cdot\vec{r}-\omega t)},\label{fields}
\end{equation}
each of which has two independent polarizations, ``plus'' and ``cross'', with allowed modes of $(\omega_+,~\omega_{\times})$, respectively. Following the spirit of DeWitt's approach \cite{DeWitt:1975ys}, the dynamics of physical gravitons in linearized gravity can be equivalent to that of two free massless scalar fields. Owing to the periodically spiral-in orbital motion, the massless scalar fields meet the periodic boundary conditions, yet on a twice separation of the original geometry. Therefore, we in turn combine the two massless scalar fields into a single massless scalar field on a twice width. Consequently, we are allowed to make gravitational analogue of the electromagnetic results and decompose the gravitoelectric and gravitomagetic fields into a scalar part and the polarization-dependent part, respectively,
\begin{eqnarray}
&&h^{\rm E}=\sum_kh^{\rm TT+}_{\rm E}\psi_k,~~h^{\rm E}=\sum_kh^{\rm TT \times}_{\rm E}\psi_k,\nonumber\\
&&h^{\rm M}=\sum_{k}h^{\rm TT+}_{\rm M}\psi_k,~~h^{\rm M}=\sum_kh^{\rm TT \times}_{\rm M}\psi_k.\label{decomfiled}
\end{eqnarray}
Here, $k$ is the wave number of the plane waves. Only the combined scalar part contributes to the gravitational Casimir effects, while the other parts form a $2\times2$ space and are responsible for the polarizations. Consequently, we just consider the dynamics that is responsible for $\psi_k$ and calculate the corresponding quantum effects in the following.

The paper is organized as follows. In section II, we tackle the planar dynamics of NS binaries that undergoes periodic boundary corrugations with the path-integral formula. Based on these formulations, we calculate the gravitational Casimir energy cost of the lateral corrugating binary plane, when the orbital separation shrinks a small distance $r$ during a nontrivial observable time. Such quantum energy cost arising from the gravitation-induced lateral corrugations of binary plane may subsequently change the orbital motion and thus the frequency of GWs during the inspiraling process. Section III considers the mechanical response of two NS components and gives qualitative investigations how the gravitational Casimir energy cost modifies orbital motion of the NS binary and impacts the frequency of released GWs. Finally, we give a brief summary for our calculations and results.

\section{Path-Integral Calculations for Gravitational Casimir Effects of Lateral Corrugating Binary Plane}

Based on the Dewitt's decompositions in Eqs. (\ref{decomfiled}), only the combined single free massless scalar field $\psi_k$ is responsible for the quantum fluctuations of gravitational field. In order to investigate the effects of quantum fluctuations, we describe the spiral-in orbital dynamics of a wide NS binary just by using the scalar action,
\begin{equation}
\mathcal{S}[\psi]=\frac{1}{2}\int d^3X\partial_{\mu}\psi(X)\partial_{\mu}\psi(X).\label{psiac}
\end{equation}
Because of the planar spiral-in orbital motion, we consider the scenario that a two-dimension binary plane embeds in a 3-dimension space-time and chose the planar polar coordinates $X^{1,2}=(r,\theta)$. The imaginary time appears as the third coordinate $X^3=ict$ due to a wick rotation. The summation over $\mu=r,\theta,t$ is automatically used in Eq. (\ref{psiac}). The field $\psi$ is naturally quantized by the constraint that it vanishes on two stars defined by $X=X_{r,\theta}(\vec{r}_{r,\theta}^{1,2})$. Accordingly, we express the partition function as \cite{Li:1991, Li:1992},
\begin{equation}
Z=\int \displaystyle\prod_{r,\theta}D\psi(\vec{r})e^{-S[\psi(\vec{r})]},
\end{equation}
and the action as
\begin{equation}
S[\psi]=\sum_{1,2}\int d\vec{r}_1d\vec{r}_2\psi(\vec{r}_1)G^3(X_1(\vec{r}_1)-X_2(\vec{r}_2))\psi(\vec{r}_2).\label{action}
\end{equation}
Here, the Green function $G^3(\vec{r})\equiv\langle\psi(\vec{r})\psi(0)\rangle_{\rm free}$ denotes the two-point correlated function of the field in free space. Integrating over the fields, we get the effective action,
\begin{equation}
S_{\rm eff}=-i\hbar \ln Z=-\frac{i\hbar}{2}\ln\mathrm{det}\{M(X_{1(2)}(\vec{r}_{1(2)}))\}.
\end{equation}
The two objects are parameterized by $(\vec{r}_1, r)$ and $(\vec{r}_2, R-r)$, respectively. $\vec{r}_1$ and $\vec{r}_2$ give the initial coordinates of two star components on the binary plane, and $r$ denotes the radial distance of orbital decay in a duration of $t$, i.e. the scale of radial deformation of the binary plane.

From Eq.(\ref{action}), we can write down the matrix as
\begin{equation}
M(\vec{r}_1,\vec{r}_2)=\left(
\begin{array}{cc}
G^3(\vec{r}_1-\vec{r}_2,0) & G^3(\vec{r}_1-\vec{r}_2,R-r) \\
G^3(\vec{r}_1-\vec{r}_2,R-r) & G^3(\vec{r}_1-\vec{r}_2,r) \\
\end{array}
\right)\label{Mmatrix}
\end{equation}
Because of the wide separation $R\sim10^9$ m and the corresponding cosmological coalescence time, the radial deformation of binary plane in an observable time (several orbital periods) can be treated as a very small quantity, i.e. $r\ll R$. As a result, we are allowed to perturbatively expand the matrix with respect to the deformations,
\begin{equation}
M(\vec{r}_1,\vec{r}_2)=M_0(\vec{r}_1,\vec{r}_2)+\delta M(\vec{r}'_1,\vec{r}'_2).\label{totalM}
\end{equation}
Here, $\vec{r}'_1,\vec{r}'_2$ are the positions of two stars after several orbital periods when the orbital decay a distance of $r$ in radial direction, whose effects are expressed as $r$ in Eq.(\ref{Mmatrix}). The matrix $M_0(\vec{r}_1,\vec{r}_2)$ in Eq.(\ref{totalM}) describes the initial state of the gravitational interaction for the binary, without any orbital decay and deformations, and reads
\begin{equation}
M_0(\vec{r}_1,\vec{r}_2)=\left(
\begin{array}{cc}
G^3(\vec{r}_1-\vec{r}_2,0) & G^3(\vec{r}_1-\vec{r}_2,R) \\
G^3(\vec{r}_1-\vec{r}_2,R) & G^3(\vec{r}_1-\vec{r}_2,0) \\
\end{array}
\right),
\end{equation}
which only depends on the relative position of two star components, i.e. the initial orbital separation $R$ of binary system. While $\delta M(\vec{r}'_1,\vec{r}'_2)$ denotes the corrections caused by the radial deformations $r$ after several orbital periods of spiral-in orbital motion. Because only the initial orbital separation $R$ and its radial shrink $r$, which describe the relative position of two stars, contribute to our calculations and results, we espress the positions $\vec{r}_1,\vec{r}_2, \vec{r}'_1,\vec{r}'_2$ as quantities describing the relative positions $R$ and $r$. Accordingly, the dependence of $\vec{r}_1,\vec{r}_2, \vec{r}'_1,\vec{r}'_2$ on time is transformed into the dependence of orbital deformation $r$ on time through the orbital period. In order to parameterized the orbital motion in our calculations by initial orbital separation $R$ and orbital shrink $r$, we diagonalize the matrix $M_0(\vec{r}_1,\vec{r}_2)$ by transforming to Fourier space, in order to describe it in terms of orbital separation only,
\begin{equation}
M_0(p,q)=\left(
\begin{array}{cc}
G^3(p) & G^3(p,R) \\
G^3(p,R) & G^3(p) \\
\end{array}
\right)(2\pi)^2\delta^2(p+q).\label{M0pq}
\end{equation}
The Fourier-transformed Green's functions are
\begin{eqnarray}
&&G^3(p)=\int G^3(\vec{r},0)e^{i\vec{p}\cdot\vec{r}}d^2\vec{r},\nonumber\\
&&G^3(p,R)=\int G^3(\vec{r},R)e^{i\vec{p}\cdot\vec{r}}d^2\vec{r},\label{Green}
\end{eqnarray}
respectively.

As a consequence, the effective action describing the spiral-in orbital motion of the binary plane consists of two parts, i.e. a gravitation-induced static action without consideration of any corrugation and the part arising from lateral spiral-in deformations,
\begin{equation}
S_{\mathrm{eff}}=S_{\mathrm{static}}+S_{\mathrm{deformation}}.
\end{equation}
The static part is given by calculating the the determinant of $M_0$,
\begin{equation}
S_{\mathrm{static}}=-\frac{i\hbar}{2}\ln\mathrm{Det}M_0.
\end{equation}
Correspondingly, the partition function of static contribution is calculated as,
\begin{equation}
\ln Z_{\mathrm{static}}=\frac{1}{2}\ln\mathrm{Det}M_0=\int\frac{d^2p}{(2\pi)^2}\ln G^3(p)+\frac{1}{2}\int\frac{d^2p}{(2\pi)^2}\ln[1-(\frac{G^3(p,R)}{G^3(p)})^2].
\end{equation}
By using the definition of Green's function and the Fourier-transformed ones, we have
\begin{equation}
G^3(p)=\frac{1}{2p},~~G^3(p,R)=\frac{e^{-pR}}{2p}.
\end{equation}
Accordingly, we obtain the contributions from static part of gravitational Casimir effects to the lateral spiral-in corrugating binary plane,
\begin{equation}
S_{\mathrm{static}}=-\frac{i\hbar}{2}\int\frac{d^2p}{(2\pi)^2}\ln\frac{1}{2p}-i\hbar\frac{\zeta(3)}{16\pi}\frac{1}{R^2}.\label{staact}
\end{equation}
In the gravitationally bound NS binary systems, the gravitational interaction and the resultant orbital motion of two objects on the binary plane make the orbital plane behave as a self-gravitation system and shrink spirally. So the first term in the static action (\ref{staact}) denotes the contribution coming from the self-gravity of the binary plane because of the periodically orbital motion. While the second term arises from the Newtonian gravitational interactions between two star components.

Due to the gravitational interactions, the two stars consisting of the binary orbit with respect to each other and move closer and closer, which leads to orbital decay and thus laterally corrugates the binary plane. The corrugations in turn induce effective interactions between two components and additional effects on the binary plane. As a result, the effective action describing the system is corrected by a deformation part, in addition to the static one, which can be computed by $\delta M$ in Eq. (\ref{totalM}),
\begin{equation}
S_{\mathrm{deformation}}=-i\hbar\ln Z_{\mathrm{deformation}}=-\frac{i\hbar}{2}\ln\mathrm{Det}(1+\frac{\delta M}{M_0}).\label{defaction}
\end{equation}
We then expand the Green's functions in Eq. (\ref{Mmatrix}) up to quadratic order,
\begin{eqnarray}
&&G^3(\vec{r}_1-\vec{r}_2, r)=G^3(\vec{r}_1-\vec{r}_2,0)+\frac{1}{2}\partial_r^2G^3(\vec{r}_1-\vec{r}_2,0)r^2,\nonumber\\
&&G^3(\vec{r}_1-\vec{r}_2, R-r)=G^3(\vec{r}_1-\vec{r}_2,R)-\partial_rG^3(\vec{r}_1-\vec{r}_2,R)r+\frac{1}{2}\partial_r^2G^3(\vec{r}_1-\vec{r}_2,R)r^2.
\end{eqnarray}
By transforming to the Fourier space, we obtain the diagonal form,
\begin{equation}
\delta M(p,q)=\left(
\begin{array}{cc}
B(p,q) & A(p,q) \\
A(p,q) & B(p,q) \\
\end{array}
\right).
\end{equation}
where
\begin{eqnarray}
A(p,q)&=&\int d^2\vec{r}_1d^2\vec{r}_2e^{i\vec{p}\cdot\vec{r}_1+i\vec{q}\cdot\vec{r}_2}\{\frac{1}{2}G^3(\vec{r}_1-\vec{r}_2,R)[(\nabla\vec{r}_1)^2+(\nabla\vec{r}_2)^2]-\partial_rG^3(\vec{r}_1-\vec{r}_2,R)r+\frac{1}{2}\partial_r^2G^3(\vec{r}_1-\vec{r}_2,R)r^2\},\nonumber\\
B(p,q)&=&\int d^2\vec{r}_1d^2\vec{r}_2e^{i\vec{p}\cdot\vec{r}_1+i\vec{q}\cdot\vec{r}_2}\{\frac{1}{2}G^3(\vec{r}_1-\vec{r}_2,0)[(\nabla\vec{r}_1)^2+(\nabla\vec{r}_2)^2]+\frac{1}{2}\partial_r^2G^3(\vec{r}_1-\vec{r}_2,0)r^2\}.
\end{eqnarray}
The inversion of matrix $M_0$ in Eq. (\ref{M0pq}) is written as
\begin{eqnarray}
M^{-1}_0(p,q)&=&\frac{1}{[G^3(p)]^2-[G^3(p,R)]^2}\left(
\begin{array}{cc}
G^3(p) & -G^3(p,R) \\
-G^3(p,R) & G^3(p) \\
\end{array}
\right)(2\pi)^2\delta^2(p+q)\nonumber\\
&=&\frac{(2p)^2}{1-e^{-2pR}}\left(
\begin{array}{cc}
G^3(p) & -G^3(p,R) \\
-G^3(p,R) & G^3(p) \\
\end{array}
\right)(2\pi)^2\delta^2(p+q).
\end{eqnarray}

Substituting the expressions of both $\delta M(p,q)$ and $M_0^{-1}(p,q)$ into the action (\ref{defaction}), we finally get the partition function induced by lateral deformations of binary plane due to the orbital decay,
\begin{equation}
\ln Z_{\mathrm{deformation}}\approx-\frac{3\zeta(3)}{16\pi R^4}\int\frac{d^2p}{(2\pi)^2}r^2+\frac{1}{16\pi^2}\int\frac{d^2p}{(2\pi)^2}r^2\frac{1}{8\pi^2(R-r)^6}.\label{defparti}
\end{equation}
The deformation-induced effective action is then written as
\begin{equation}
S_{\mathrm{deformation}}\approx i\hbar\frac{3\zeta(3)}{16\pi R^4}\int\frac{d^2p}{(2\pi)^2}r^2-\frac{i\hbar}{16\pi^2}\int\frac{d^2p}{(2\pi)^2}r^2\frac{1}{8\pi^2(R-r)^6}.\label{defpart}
\end{equation}
The first term ranges as $\sim\frac{1}{(R-r)^2}$ and represents the perturbations of the Newtonian contributions to the gravitational Casimir energy, arising from the instability of orbital binding energy induced by the spiral-in orbital motions. The second term, decaying as $\sim\frac{1}{(R-r)^4}$, gives the gravitational Casimir energy cost of the binary plane that comes from the gravitational-controlled orbital decay, i.e. the lateral spiral-in deformations or corrugations of orbital plane.

\section{laterally Corrugating Binary Plane}

The gravitational Casimir effects on the binary plane during the inspiraling process of NS binaries, expressed by Eq. (\ref{staact}) and Eq. (\ref{defpart}), may modify the dynamics described by Einstein's general relativity. In order to study the corrections to the dynamics of NS binary system due to the gravitational Casimir energy cost resulting from lateral corrugations of binary plane, we diagonalize the expression of deformation action (\ref{defpart}) by using Fourier transformations. The radial shrinks can be transformed as,
\begin{equation}
r(\vec{r},t)=\int\frac{d\omega d^2\vec{k}}{(2\pi)^3}e^{-i\omega t+i\vec{k}\cdot{\vec{r}}}r(\vec{k},\omega),
\end{equation}
where we use the definition $r(t)=\int d\omega r(\omega)e^{-i\omega t}$ for the time-dependent part.
Because of the time dependency of lateral corrugations of the orbital plane, we perform a rotation from imaginary time $it$ to a real one $\tau$. Accordingly, the diagonal deformation part of effective action is written as
\begin{equation}
S_{\mathrm{deformation}}=\frac{\hbar c}{2}\int\frac{d\omega d^2q}{(2\pi)^3}[r^2(k,\omega)A_+(k,\omega)-A_-(k,\omega)r(k,\omega)r(-k,-\omega)].
\end{equation}
Here, $A_{\pm}(k,\omega)$ are the kernels and represent a close relation to the mechanical response of NS binary plane system due to the gravitational interactions between two star components, which are functions of the binary separation $R$ and the distance of radial decay $r$.
\begin{eqnarray}
A_+(k,\omega)&=&\iint d\tau d^2re^{-\omega\tau+i\vec{k}\cdot\vec{r}}\big\{\partial_r^2G(\sqrt{r^2+\tau^2},0)\int\frac{d^2p}{(2\pi)^2}\frac{G^3(p)}{[G^3(p)]^2-[G^3(p,R)]^2}e^{i\vec{p}\cdot\vec{r}}\nonumber\\
&&+\int\frac{d^2p}{(2\pi)^2}\frac{G^3(p)}{[G^3(p)]^2-[G^3(p,R)]^2}e^{i\vec{p}\cdot\vec{r}}\cdot\int\frac{d^2p}{(2\pi)^2}\frac{G^3(p)}{[G^3(p)]^2-[G^3(p,R)]^2}(\frac{\partial G^3(p,R)}{\partial R})^2e^{i\vec{p}\cdot\vec{r}}\nonumber\\
&&+\int\frac{d^2p}{(2\pi)^2}\frac{G^3(p,R)}{[G^3(p)]^2-[G^3(p,R)]^2}e^{i\vec{p}\cdot\vec{r}}\cdot\int\frac{d^2p}{(2\pi)^2}\frac{G^3(p,R)}{[G^3(p)]^2-[G^3(p,R)]^2}(\frac{\partial G^3(p,R)}{\partial R})^2e^{i\vec{p}\cdot\vec{r}}\big\},\\
A_-(k,\omega)&=&\iint d\tau d^2re^{-\omega\tau+i\vec{k}\cdot\vec{r}}\big\{\partial_r^2G(\sqrt{r^2+\tau^2},R)\int\frac{d^2p}{(2\pi)^2}\frac{G^3(p,R)}{[G^3(p)]^2-[G^3(p,R)]^2}e^{i\vec{p}\cdot\vec{r}}\nonumber\\
&&+\big(\int\frac{d^2p}{(2\pi)^2}\frac{G^3(p)}{[G^3(p)]^2-[G^3(p,R)]^2}\frac{\partial G^3(p,R)}{\partial R}e^{i\vec{p}\cdot\vec{r}}\big)^2\nonumber\\
&&+\big(\int\frac{d^2p}{(2\pi)^2}\frac{G^3(p,R)}{[G^3(p)]^2-[G^3(p,R)]^2}\frac{\partial G^3(p,R)}{\partial R}e^{i\vec{p}\cdot\vec{r}}\big)^2\big\}.
\end{eqnarray}
The exact forms of these kernels are not illuminating and need cucumber integrals. However, their behaviors depend on $k$ and $\omega$ only through the relation $\omega^2-c^2k^2$, and the competition of $\omega<ck$ and $\omega>ck$ displays different scenarios \cite{Golestanian:1997ks}, which provides enlightening approximative results for a qualitative analysis in particular systems.

By taking the scales of $\sim10^4$ m for an NS and the cosmological time for coalescence into consideration, the separation of $R\sim10^9$ m in wide inspiraling NS binary systems can be treated as $R\to\infty$. In the limit of separation $R\to\infty$, the kernel $A_-^{\infty}(k,\omega)$ disappears, i.e. $A_-^{\infty}(k,\omega)=0$. According to Einstein's general relativity, GWs propagate at the speed of light, which implies that the graviton is massless and obeys the dispersion relation of $\omega=ck$. However, the inspiraling NS binary system involves dissipation of orbital binding energy, which is characterized by the orbital decay and the continual release of GWs (or generation of gravitons). It is a purely imaginary mechanical response function that signifies the dissipation of energy \cite{MaiaNeto:1993zz} and the generation of corresponding quanta \cite{Lambrecht:1996un}, i.e. gravitons in inspiraling NS binaries, which is represent by a purely imaginary kernel $A^{\infty}_+(\omega)$ in flat background,
\begin{equation}
A_+^{\infty}(\omega)\approx i\frac{\mathrm{sgn}(\omega)\omega^5}{360\pi^2c^5}.\label{aplus}
\end{equation}
$\mathrm{sgn}(\omega)$ therein is the sign function. Consequently, both of the moving stars undergo an additionally dissipative radiation reaction force, in addition to the gravitational Casimir interaction and the Newtonian gravitational interaction.

According to the linear response theory \cite{Kubo1966}, the response to a given perturbation, i.e. gravitation-induced spiral-in motion, is described by a gravitational susceptibility function $\chi(t)$. When the system experiences a small displacement $r(t)$, the field reacts by exerting a force along the direction of the displacement,
\begin{equation}
F(t)=\int dt\chi(t)r(t),
\end{equation}
which is similar to the frictional force of Eq. (\ref{friction}) imposed by slowly moving boundaries. As a consequence, the laterally corrugating binary plane is exerted a lateral friction-like force in the radial direction. By considering the frequency of the emitted GWs increasing due to the radial displacement during the inspiraling process. Such a radial force depends on the frequency of GWs,
\begin{equation}
F(\omega)=\frac{\partial S_{\mathrm{deformation}}}{\partial r}=\chi(\omega)r(\omega).
\end{equation}
In the inspiraling NS binaries due to the gravitational interaction of two orbiting star components, the gravitational susceptibility represents the mechanical response tensor of the system, which reads,
\begin{equation}
\chi(\omega)=c\hbar\int\frac{d^2k}{(2\pi)^2}k^2[A_+(k,\omega)-A_+(k,\omega')]r^2.
\end{equation}
So the imaginary part of the gravitational susceptibility $\chi(\omega)$ corresponds to the dissipative force $F(\omega)$ and signals a dissipation energy cost from the binary plane. In order to study the influences on the frequency of GWs, we, accordingly, introduce an effective coefficient of viscosity, defined by
\begin{equation}
\chi(\omega)=-i\omega\eta(\omega),\label{chiomega}
\end{equation}
where $\eta(\omega)$ is frequency-dependent effective viscosity. Because of the radial dependency of the dissipative force, $\eta(\omega)$ should be anisotropic and just shows the radial component, which is estimated as
\begin{equation}
\eta_r(\omega)=\frac{\hbar Rr^2\omega^4}{1440\pi c^4\lambda_g^2}.\label{vis}
\end{equation}
The other components of $\eta(\omega)$ disappear.

Suffering from the dissipation force, the dissipation rate of the orbital bing energy can be calculated as
\begin{equation}
P=\frac{1}{T}\int_0^Tdt\dot{r}(t)F(t)=-\frac{1}{T}\int\frac{d\omega}{2\pi}\omega\chi(\omega)r^2(\omega),\label{rate}
\end{equation}
where $T$ denotes one orbital period. Substituting Eq. (\ref{vis}) and Eq. (\ref{chiomega}) into Eq. (\ref{rate}), we can find that such an dissipative cost of the orbital binding energy increases with the orbital decay and the increasing GWs frequency during the spiral-in process, which subsequently accelerates the orbital decay of the binary. However, the dissipation is proportional to six times power of GW frequencies. By considering that the GWs from the wide systems have the characteristic frequencies of $10^{-3}-10^{-4}$ Hz, the dissipation rate has an order of $10^{-70}$ eV/s. Therefore, the effects of acceleration on the orbital decay owing to the dissipation force is very small.

\section{Summary}

As a summary, we use the path-integral approach and calculate the gravitational Casimir effects induced by the lateral corrugation of orbital plane due to the spiral-in motion of the inspiraling NS binary. Based on Dewitt's approach, we decompose the spin-2 gravitons into the polarization parts and a scalar part that is responsible for the quantum effects. The effects due to quantum fluctuations of the gravitational field induced by the lateral corrugations of the binary plane include a static part and a deformation-induced term. It is found that, without considering the corrugation, the contributions to the gravitational Casimir effects consist of the self-gravitation of the binary plane and the Newtonian gravitational interactions of the binary system ranging as $\frac{1}{R^2}$. Previously, the leading-order quantum corrections to Newtonian gravitational interaction between two massive and stationary objects have been computed \cite{Donoghue:1994dn}. The quantum contributions of Eq.(\ref{staact}) from the Newtonian gravitational interactions, i.e. the static gravitational Casimir effects, gives a supplement to the result of \cite{Donoghue:1994dn}, i.e. quantum corrections to the Newtonian gravitational interaction, at order $\mathcal{O}(h)$ in massive gravitational bound systems when considering their motions, such as orbiting NS binaries. Because of the gravitational interactions, the two NS components move closer and closer on the binary plane in a spiral-in way, which cause the boundary of the orbital plane continually corrugate in a periodically spiral-in way, i.e. the lateral deformation of the binary plane. Such a corrugation allows us to perturbatively study the gravitational Casimir effects in the gravitationally controlled dynamics of binaries. The spiral-in deformation of the orbital plane leads to instability of orbital binding energy, which scales as $\frac{1}{(R-r)^2}$. While the gravitational induced orbital decay contributes to additional quantum cost of the orbital binding energy, which decays as $\frac{1}{(R-r)^4}$. Both of these two kinds of quantum energy cost are responsible for the dynamically gravitational Casimir energy cost induced by the laterally spiral-in deformation of the binary plane.

The dynamically gravitational Casimir phenomena due to slowly moving boundaries associate with production of gravitons. The binary system suffers from an additional energy cost, because of energy conservation. Accordingly, the dynamical Casimir effects exert a dissipative force on the binary plane along the direction of the displacement, i.e. a radial force. As a result, the spiral-in orbital decay of the binary and the frequency of the released GWs suffer from some quantum corrections to the dynamics controlled by Einstein's general relativity. In order to investigate the corrections, we consider the mechanical response of two star components and its dependence on the frequency of released GWs, and calculate the dissipation effects qualitatively. An anisotropic characteristic viscosity coefficient, just shows the radial component, represent the mechanical response of the binary system, which is the function of the scales of radial shrink and the frequency of GWs. Consequently, the dissipation rate of the orbital binding energy may increase with the decaying separation and increasing GW frequencies, which will subsequently accelerates the orbital decay of the binary. However, the GWs from spiral-in orbital decay of wide NS binaries with separation of $10^9$m locate in the low-frequency band of $10^{-3}-10^{-4}$ Hz. The dissipation rate just has an order of $10^{-70}$ eV/s. Therefore, the accelerating effects on the orbital decay due to the additional quantum energy cost is very marginal.

\section*{Acknowledgements}
This work is supported by the Guangxi Natural Science Foundation Program (Grant no. 007151339018) and by Guangxi Science and Technology Foundation and Talent Special (Grant no. 111252047014).

\end{document}